\documentclass[pre,preprintnumbers,amsmath,amssymb]{revtex4}
\usepackage{graphicx}% Include figure files
\usepackage{bm}% bold math

\begin{document}
\title{Real-space Condensation in Stochastic Mass Transport Models}
\author {Satya N. Majumdar }

\affiliation{ Laboratoire de Physique Th\'eorique et Mod\`eles Statistiques,
        Universit\'e Paris-Sud. B\^at. 100. 91405 Orsay Cedex. France}

\begin{abstract}
The phenomenon of real-space condensation is
encountered in a variety of situations such as aggregation and fragmentation processes,
granular clustering, phase separation, traffic and networks.
Unlike traditional Bose-Einstein condensation in the {\it momentum space},
a condensate in these systems forms in {\it real space}, e.g., upon
increasing the density beyond a critical value a macroscopically large
mass/cluster may form at a single site on a lattice. 
In this brief review, I discuss some recent developments in 
understanding the physical and mathematical mechanism behind
this real-space condensation in a class of simple stochastic mass
transport models.

\end{abstract}

\maketitle

\date{\today}

\section{introduction}

The phenomenon of Bose-Einstein condensation (BEC) in an ideal Bose gas is by now a textbook 
material
and has recently seen a huge revival of interest driven mostly by new experiments. Consider 
an ideal gas of $N$ bosons in a $d$-dimensional hybercubic box of volume $V=L^d$.
In the thermodynamic limit $N\to \infty$, $V\to \infty$ but with the density
$\rho= N/V$ fixed, as one reduces the temperature below a certain
critical value $T_{c}(\rho)$ in $d>2$, macroscopically large number of
particles ($\propto V$) condense on to the ground state, i.e., in the zero momentum
quantum state. Alternately, one encounters the same condensation transition upon
fixing the temperature but increasing the density $\rho$ beyond a critical
value $\rho_c(T)$. 

The traditional BEC happens in the momentum
(or equivalently energy) space. In contrast, over the last two decades it has been realized that
a `similar' Bose-Einstein type condensation  
can also occur in {\it real space} in the steady state of a variety
of nonequilibrium systems such as in cluster aggregation and 
fragmentation~\cite{MKB}, jamming in 
traffic
and granular flow~\cite{OEC,CSS} and granular clustering~\cite{granular}.
The common characteristic feature 
that these systems share is the stochastic transport of some conserved scalar quantity
which can simply be called mass. Condensation 
transition in these systems occurs when above some critical mass density a single `condensate' 
captures a finite fraction of the total mass of the system. `Condensate' corresponds
to a dominant cluster in the context of granular clustering, or a single large jam in 
the context of traffic models. Another example of condensation is found
in the phase separation dynamics in one dimensional driven systems where
the condensation manifests itself in the emergence of a macroscopic domain of
one phase~\cite{phase-separation}. Other examples of 
such 
real-space condensation
can be found in the socioeconomic contexts: for example, in wealth condensation in
macroeconomics where a single individual or a company (condensate) owns a
finite fraction of the total wealth~\cite{wealth} or in growing networks where a single node or hub
(such as `google') may capture a finite fraction of links in the network~\cite{networks}.

This real-space condensation mentioned above
has been studied theoretically in very simple stochastic mass transport 
models defined on lattices.
These are typically nonequilibrium models without any Hamiltonian and are defined
by their microscopic dynamical rules that specify how some scalar quantities such as
masses or a certain number of particles get transported from site to site of the lattice. 
These rules typically violate detailed balance~\cite{evans_review}.
Under these rules, 
the system evolves into
a {\it stationary} or {\it steady} state which are typically not Gibbs-Boltzmann
state as the system lacks a Hamiltonian~\cite{evans_review}. For a class of transport rules, 
the system can reach a steady state where upon increasing the density
of mass or particles beyond a critical value, a macroscopically
large mass (or number of particles) condenses onto a single site of the lattice, signalling
the onset of `real-space' condensation. 

In this article we will mostly
focus on {\it homogeneous} systems where the transport rules are independent of sites, i.e.,
the system is translationally invariant. In the condensed phase and in
an infinite system, the condensate forms
at a single site which thus breaks the translational invariance spontaneously.
In a finite system, the condensate at a given site has a finite lifetime beyond
which it dissolves and then gets relocated at a different site and the various 
time scales associated with the formation/relocation of the condensate
diverge with the increasing system size (see later). 
In {\it heterogeneous} systems where the transport rules may differ from site to site, 
the 
condensate may form at a site with the lowest outgoing mass transport rate~\cite{Evans1,KF,AEM}.
The mechanism of the condensation transition in such heterogeneous systems 
is exactly analogus to the traditional BEC in
momentum space and the site with the lowest outgoing mass trasfer rate
plays the role of the ground state in the quantum system of ideal Bose gas.
In contrast, the mechanism of condensation in homogeneous systems, 
the subject of focus here, is rather different: the onset and formation of a condensate
in an infinite system is associated with the spontaneous breaking of translational 
invariance.
Also, unlike the traditional equilibrium Bose gas in a box, this real-space condensation
in nonequilibrium mass transport models can occur {\it even
in one dimension}.

The purpose of these lectures is to understand the phenomenon of real-space condensation 
in homogeneous systems within the context of 
simple one dimensional
mass transport models. The main questions
we will be addressing are threefold: (i) When does the condensation happen-- i.e. to find
the criterion
for condensation (ii) How does the condensation happen--i.e., to unfold the mathematical mechanism
behind such a transition if it happens and (iii) What is the nature of the condensate-- e.g.,
to compute the distribution of mass or the number of particles in the condensate.

The article is organized as follows. In Section II we will discuss three simple
and well studied lattice models of stochastic mass transport. In Section III, we
will consider a generalized mass transport model that includes the previous three
models as special cases and investigate its steady state. In particular, we
will study in detail steady states that are factorisable. The necessary
and sufficient conditions for such factorisable property will be discussed.
Thanks to this factorisable property, a detailed analytical study of the
condensation is possible for such steady states which will be illustrated in
Section IV. In Section V we will illustrate how various results associated with
the condensation transition in factorisable steady states can be simply understood
in terms of sums and extremes of independent and identically distributed (i.i.d) 
random variables. Finally we will conclude in 
Section VI with a summary and other
possible generalizations/issues associated with the real-space condensation.

\section{Three Simple Mass Transport Models}

\subsection{Zero Range Process}

The Zero Range Process (ZRP), introduced by Spitzer~\cite{Spitzer}, is perhaps one of the 
simplest 
analytically solvable
model of mass/particle transport that exhibits a real-space condensation in
certain range of its parameters--for a review see ~\cite{zrp_review}. ZRP is defined on a 
lattice with
periodic boundary conditions. For simplicity,
we will consider a $1$-d lattice with $L$ sites, the generalization to higher dimensions is
straightforward. On each site of the lattice at any instant of time rests
a number of particles, say $m_i$ at site $i$ where $m_i\ge 0$ is a
nonnegative integer. We can also think of each particle carrying a unit mass, so
that $m_i$ represents the total mass at site $i$. A configuration of the 
system at any given instant is specified by 
the masses at all sites $\{m_1,m_2,\ldots ,m_L\}$. One starts
from an arbitrary initial condition with a total mass $M=\sum_i m_i$.
The subsequent dynamics conserves this total mass or total particle number, or equivalently the
density $\rho=M/L$.

The system evolves via continuous-time stochastic dynamics specified by the following rules:

\vspace{0.2cm}

$\bullet$ In a small time interval $dt$, a single particle from site $i$ with $m_i$ number
of particles is transported to its right neighbour $i+1$ with probability $U(m_i) dt$
provided $m_i\ge 1$. In terms of mass, this means a single unit of mass is transferred from 
site
$i$ to site $i+1$ with rate $U(m_i)$.

\vspace{0.2cm}

$\bullet$ Nothing happens with probability $1-U(m_i)dt$.

\vspace{0.2cm}

\noindent Here $U(m)$ is an arbitrary positive function, with the constraint
that $U(0)=0$, since there can not be any transfer of unit mass
if the site has no mass at all. Thus in ZRP, the 
particle or mass
transfer rate $U(m)$ depends only on the number of particles/mass $m$ at the departure site 
prior
to the transfer. One can of course generalize easily the ZRP to  
discrete-time dynamics, with
symmetric transfer of particles to both neighbours etc ~\cite{zrp_review}. But here we stick 
to the
asymmetric continuous-time model for simplicity. 

As the system evolves under this dynamics, the probability of a configuration
$P(m_1,m_2,\ldots,m_L,t)$ evolves in time and in the long time limit, $t\to \infty$, it approaches
a time-independent stationary joint distribution of masses
$P(m_1,m_2,\ldots, m_L)$. This is the basic quantity of interest, since
the statistics of all other physical observables in the steady state can, in principle, be computed from 
this joint distribution. In many such nonequilibrium systems, computing the steady state
$P(m_1,m_2,\ldots,m_L)$ is, indeed, the first big hurdle~\cite{evans_review}. Fortunately, in ZRP,
this can be computed explicitly and has a rather simple factorised form~\cite{Spitzer,zrp_review}  
\begin{equation}
P(m_1,m_2,\ldots,m_L)= \frac{1}{Z_L(M)}\, f(m_1)f(m_2)\ldots f(m_L)\, \delta\left(\sum_{i} m_i -M\right)
\label{zrp_jpm1}
\end{equation}
where the weight function $f(m)$ is related to the transfer rate $U(m)$ via
\begin{eqnarray}
f(m)&= &\prod_{k=1}^m \frac{1}{U(k)} \quad\, {\rm for}\quad m\ge 1 \nonumber \\
&=& 1 \quad\quad {\rm for} \quad m=0
\label{zrp_weight1}
\end{eqnarray}
The delta function in Eq. (\ref{zrp_jpm1}) specifies the conserved total mass $M$
and $Z_L(M)$ is just a normalization factor that ensures that the total probability is unity
and satisfies a simple recursion relation
\begin{equation}
Z_L(M)= \sum_{m_i} \prod_{i=1}^L f(m_i) \, \delta\left(\sum_i m_i-M\right)= \sum_{m=0}^M f(m) 
Z_{L-1}(M-m).
\label{norm_zrp1}
\end{equation}
To prove the result in Eq. (\ref{zrp_jpm1}) one simply writes down the Master equation
for the evolution of the probability in the configuration space and then verifies~\cite{zrp_review}
that the stationary solution of this Master equation is indeed given by (\ref{zrp_jpm1}).

Finally, the single site mass distribution $p(m)$, defined as the probability
that any site has mass $m$ in the steady state, is just the marginal
obtained from the joint distribution 
\begin{equation}
p(m) = \sum_{m_2,m_3,\ldots, m_L} P(m,m_2,m_3,\ldots, m_L)= f(m)\, 
\frac{Z_{L-1}(M-m)}{Z_L(m)}.
\label{zrp_spm1}
\end{equation}
Note that $p(m)$ implicitly depends on $L$ but this $L$ dependence has been suppressed
for notational simplicity.
This single site mass distribution is important as any signature of the existence of a
condensate will definitely show up in the explicit form of $p(m)$.

Evidently the steady state $p(m)$ depends on the transfer rate $U(m)$ through the
weight function $f(m)$ in Eq. (\ref{zrp_weight1}). 
Not all choices of $U(m)$ lead
to a steady state with a condensation transition. Indeed, one may ask what choices of $U(m)$ may lead 
to a condensation transition. An example of such a choice is given by
$U(m) \propto (1+\gamma/m)$ for large $m$, which leads to, using Eq. (\ref{zrp_weight1}),
a power law weight function, $f(m)\sim m^{-\gamma}$ for large $m$. In this case,
it was shown~\cite{OEC,zrp_review} that for $\gamma>2$, the system undergoes a condensation transition
as one increases the density $\rho$ through a critical value $\rho_c=1/(\gamma-2)$.
The condensation transition shows up in $p(m)$ in the thermodynamic limit, which
has different forms for $\rho<\rho_c$, $\rho=\rho_c$ and $\rho>\rho_c$~\cite{zrp_review}
\begin{eqnarray}
p(m) & \sim & \frac{1}{m^{\gamma}}\, \exp\left[-m/m^*\right]\quad {\rm for} \quad \rho<\rho_c 
\nonumber \\
&\sim &\frac{1}{m^{\gamma}} \quad {\rm for} \quad \rho=\rho_c
\nonumber \\
&\sim & \frac{1}{m^{\gamma}} + {\rm "condensate"} \quad {\rm for} \quad \rho>\rho_c
\label{zrp_spm2}
\end{eqnarray}
Thus for $\rho<\rho_c$, the single site mass distribution decays exponentially with a characteristic 
mass $m^*$ that diverges as $\rho\to \rho_c$ from below, has a power law form exactly at $\rho=\rho_c$
and for $\rho>\rho_c$, while the power-law form remains unchanged, all the additional mass 
$(\rho-\rho_c)L$ condenses onto a single site which shows up as a bump in $p(m)$ at the tail
of the power law form (see Fig. (\ref{fig:dist})). The term ``condensate" in
Eq. (\ref{zrp_spm2}) refers to this additional bump. Physically this 
means that a single condensate coexists with
a background critical fluid for $\rho>\rho_c$.
This change of behavior of $p(m)$ as one increases $\rho$ through $\rho_c$ is a prototype
signature of the real-space condensation and one finds this behavior
in various other 
stochastic mass
transport models that will be discussed below. In addition, many details of the condensation phenomena in 
ZRP
also follow as special cases of the more general mass transport model
defined in Section III.
 
Before ending this subsection, it is useful to 
point out that there have been several other issues and studies on ZRP and related models that
are not covered here. The interested readers may consult the reviews~\cite{zrp_review,godreche_review}.
  
\subsection{Symmetric Chipping Model}

Here we discuss another simple one dimensional mass transport model that also exhibits
a condensation phase transition in its steady state. As in ZRP, this model is also defined 
on a lattice with periodic boundary conditions where each site $i$ carries an integer
mass $m_i\ge 0$~\cite{MKB}. A non-lattice mean-field version of the model was  
studied in Ref.~\cite{KrapRed}.
The system evolves via the continuous-time dynamics defined by the 
following rules~\cite{MKB,MKBlong}:

\vspace{0.2cm}

$\bullet$ {\it diffusion and aggregation}: in a small time interval $dt$, the entire mass 
$m_i$ from site 
$i$ moves
either to its right neighbour $(i+1)$ with probability $dt/2$ or to its left neighbour
$(i-1)$ with probability $dt/2$. 
%After the transfer, say to site $(i+1)$, the total
%mass at $(i+1)$ just adds up , i.e., $m_{i+1}\to m_{i+1}+m_i$.

\vspace{0.2cm}

$\bullet$ {\it chipping}: in the same interval $dt$, only one unit of mass chips off site $i$ with mass
$m_i$ (provided $m_i\ge 1$) to either 
its right neighbour with probability $w\,dt/2$ or to its left neighbour with probability
$w\,dt/2$. 

\vspace{0.2cm} 

$\bullet$ With probability $1-(1+w)\,dt$ nothing happens.

\vspace{0.2cm}

\noindent Once again the total mass $M=\rho L$ is conserved by the dynamics. The model thus has two 
parameters
$\rho$ (density)and $w$ (the ratio of the chipping to the diffusion rate). At long times, 
the
system evolves into a steady state where the single site mass distribution $p(m)$, for large $L$,
exhibits a condensation phase transition at a critical density~\cite{MKB}, $\rho_c(w)= 
\sqrt{w+1}-1$. Remarkably, this equation of state $\rho_c(w)=\sqrt{w+1}-1$ turns
out to be exact in all dimensions~\cite{RMchipping} and is thus `superuniversal'.
For $\rho<\rho_c(w)$, the mass is homogeneously distributed in the system with a
mass distribution that has an exponential tail for large mass.
At $\rho=\rho_c(w)$, the mass distribution decays 
as a power law and for $\rho>\rho_c$, a condensate forms on a single site that carries
the additional macroscopic mass $(\rho-\rho_c)L$ and coexists with a critical
background fluid~\cite{MKB}
\begin{eqnarray}
p(m) & \sim & \exp\left[-m/m^*\right]\quad {\rm for} \quad \rho<\rho_c(w)
\nonumber \\
&\sim &\frac{1}{m^{\tau}} \quad {\rm for} \quad \rho=\rho_c(w)
\nonumber \\
&\sim & \frac{1}{m^{\tau}} + {\rm "condensate"} \quad {\rm for} \quad \rho>\rho_c
\label{chipping_spm2}
\end{eqnarray}
where the exponent $\tau=5/2$ within the mean field theory~\cite{MKB} and is conjectured
to have the same mean field value even in one dimension~\cite{RMchipping}. 

Note that unlike ZRP, the exact joint distribution of masses
$P(m_1,m_2,\ldots,m_L)$ in the steady state is not known for the symmetric chipping model.
In fact, it is believed~\cite{RMchipping} that $P(m_1,m_2,\ldots,m_L)$ does not have a simple
product measure (factorisable) form as in ZRP in Eq. (\ref{zrp_jpm1}).          
Another important difference is that in ZRP, the condensation transition happens
both for asymmetric as well as symmetric transfer of masses to the neighbours,
as long as one chooses the rate $U(m)$ appropriately.
In contrast, for the chipping model, a true condensation transition happens in
the thermodynamic limit only for the symmetric transfer of masses. For the
asymmetric transfer of masses (say only to the right neighbour), the condensed phase
disapears in the thermodynamic limit even though for finite $L$ one does see
a vestige of condensation transition~\cite{RKchipping}. However, a generalization that includes both the 
chipping model and ZRP as special cases does appear to have a condensation transition even
with asymmetric hopping~\cite{LMZ}. Finally, when the diffusion rate depends on
the mass of the departure site in the chipping model in a certain manner, the condensation transition
disappears~\cite{RDBB,LKK}.

This simple chipping model with aggregation and fragmentation rules have 
been useful in various experimental contexts such as in the growth of 
palladium nanoparticles~\cite{VA1}. Besides, the possibility of
such a condensation phase transition driven by the aggregation mechanism has been discussed 
in a system of Au sputtered by swift heavy ions~\cite{Kuiri}.
Finally, the chipping model and its various generalizations have also been studied 
in the context of traffic~\cite{LZG}, finance~\cite{chipping_finance} and 
networks~\cite{chipping_network}.

\subsection{Asymmetric Random Average Process}

Another simple mass transport model that has been studied 
extensively~\cite{ARAP1,ARAP2,RMARAP,ZS} is the
asymmetric random average process (ARAP). 
As in the previous two models,
ARAP is defined on a one dimensional lattice with periodic boundary conditions.
However, in contrast to ZRP 
and 
the chipping model,
here the mass $m_i$ at each site $i$ is assumed to be a continuous positive variable.
The model has been studied for both continuous-time as well as discrete-time 
dynamics~\cite{ARAP1,ARAP2}.
In the continuous-time version, the microscopic evolution rules are~\cite{ARAP1,ARAP2}:

\vspace{0.2cm}

$\bullet$ In a small time interval $dt$, a random fraction $r_i\,m_i$ of the mass $m_i$
at site $i$ is transported to the right neighbour $(i+1)$ with probability $dt$, where
$r_i\in [0,1]$ is a random number chosen, independently for each site $i$, from 
a uniform distribution over $[0,1]$.

\vspace{0.2cm}

$\bullet$ With probability $1-dt$, nothing happens.

\vspace{0.2cm}
  
\noindent The dynamics evidently conserves the total mass $M=\rho L$. At long times,
the system reaches a steady state. Once again, the joint distribution of masses
$P(m_1,m_2,\ldots,m_L)$ 
in the steady state does not have a factorised product measure form as in ZRP.

How does the single site mass distribution $p(m)$ look like in the large $L$ limit?
First important point one notices here is that the density $\rho$ obviously sets the overall mass
scale in this model. In other words, the mass distribution $p(m,\rho)$
for any given $\rho$
must have an exact scaling form
\begin{equation}
p(m,\rho) = \frac{1}{\rho} F\left(\frac{m}{\rho}\right)
\label{arap_sc1}
\end{equation}
where the scaling function $F(x)$ must satisfy conditions
\begin{equation}
\int_0^{\infty} F(x)dx=1; \quad \quad \int_0^{\infty} x\, F(x)\, dx =1.
\label{arap_norm1}
\end{equation}
The first condition follows from normalization, $\int p(m,\rho) dm=1$ and the
second from the mass conservation, $\int m\, p(m,\rho)\, dm = \rho$.
Since the dynamics involves transfering a uniform fraction of mass from one
site to its neighbour, the scaling in Eq. (\ref{arap_sc1}) is preserved by
the dynamics. This is in contrast to ZRP or the chipping model, where
one chips of a single unit of mass from one site to its neighbour and thereby
the dynamics introduces a separate mass scale (unit mass), in addition
to the overall density $\rho$.
  
The scaling function $F(x)$ for ARAP has been computed within the mean field theory~\cite{ARAP1,ARAP2}
\begin{equation}
F(x) = \frac{1}{\sqrt{2\pi x}}\, e^{-x/2}
\label{arap_mf1}
\end{equation}
and this mean field result is remarkably close to the numerical results in one dimension,
even though one can prove rigorously~\cite{ARAP2} that the joint distribution of masses do not factorise.
In contrast,
for ARAP defined with a parallel discrete-time dynamics (where all sites are updated
simultaneously), it has been proved~\cite{ARAP1,ARAP2} that the joint distribution of masses 
factorise
as in ZRP and the scaling function $F(x)$ for the single site mass distribution
can be computed exactly, $F(x)= 4\,x\,e^{-2x}$. The steady state of the discrete-time
ARAP is also related to the steady state of the so called $q$-model of force fluctuations
in granular materials~\cite{qmodel}

What about condensation? In ARAP, one does not find a condensation transition.
This is of course expected since the density $\rho$ just sets the mass scale
and one does not expect to see a change of behavior in the mass distribution
upon increasing $\rho$, apart from a trivial rescaling of the mass at all sites
by a constant factor $\rho$. However, one can induce a condensation transition
in ARAP by inducing an additional mass scale, e.g., by imposing
a maximum threshold on the amount of mass that may be transferred from
a site to its neighbour~\cite{ZS}.

\section{A Generalized Mass Transport Model}

Let us reflect for a moment what we have learnt so far from the three models discussed
above. It is clear that the dynamics of mass transport often, though not always,
may lead to a steady state that exhibits real-space condensation. For example,
the ZRP and the symmetric chipping model exhibit real-space condensation, but
not the ARAP. Also, we note that some of these models such as ZRP have a simple
factorisable steady state. But {\it factorisability 
of the steady state is clearly not a necessary condition for a system to exhibit real-space
condensation}, as we have learnt, e.g., from the study of the symmetric chipping model
where the steady state is not factorisable.
The factorisability, if present, of course helps the mathematical analysis.

So, a natural question is: given a set of microscopic mass transport rules, what
are the necessary and sufficient conditions that they may lead to a steady state that 
exhibits real-space condensation? For example, from the study of the three models above
it seems that in order to have a condensation one needs to introduce via
the dynamics a different mass 
scale, in addition to the density, such as the chipping of a single unit 
of mass in ZRP
and the symmetric chipping model, or via introducing a maximum cap on
the mass to be transferred in ARAP. If there is only one 
overall mass scale (density) that
is preserved by the dynamics as in the usual ARAP, one does not expect to see a phase
transition in the mass distribution as one changes the density.   

This question about finding the conditions for real-space condensation in
a generic mass transport model seems too general and is perhaps difficult
to answer. Instead, one useful strategy is to restrict ourselves to 
a special class of mass transport models that have a factorisable
steady state and then ask for the criterion, mechanism and nature
of the real-space condensation phenomenon within this restricted
class of mass transport models, which includes ZRP as a special class.
This strategy has been demonstrated to work rather
successfully in a recent series of papers~\cite{EMZ1,ZEM,MEZ,EMZlong,EMZgraph} 
and a fairly good understanding of the real-space
condensation phenonemon has been developed within this restricted class of mass transport 
models. This is what we will briefly discuss in this section.

\vspace{0.2cm}

\begin{figure}[tbph]
\includegraphics[height=10.0cm,width=10.0cm,angle=0]{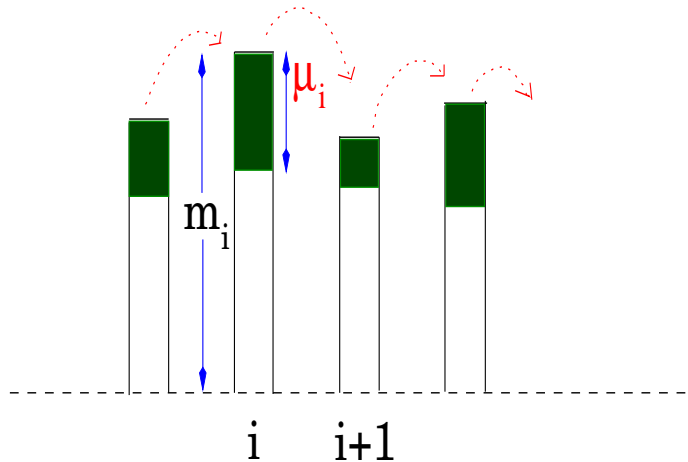}
\caption{Generalized mass transport model where a random mass $\mu_i$
is transfered from site $i$ with mass $m_i$ to site $i+1$.}
\label{fig:transport}
\end{figure}

{\bf A Generalized Mass Transport Model:} One can include all the three
models discussed in Section II in a more generalized mass transport model~\cite{EMZ1}.
For simplicity we define the model here on a one dimensional ring of $L$ sites with 
asymmetric
transfer rules, but it can be generalized in a straightforward manner on arbitrary graphs
and arbitrary transfer rules. Similar mass transport models with open
boundaries as well as dissipation at each site
have also been studied~\cite{Bertin}, though here we restrict ourselves to
periodic boundary conditions and non-dissipative dynamics that preserve the total mass. 
On each site of the ring there is a scalar continuous mass 
$m_i$.
At any given time $t$  one chooses
a mass $0\le \mu_i\le m_i$ independently at each site from a probability
distribution $\phi(\mu_i|m_i)$, normalized such that $\int_0^m \phi(\mu|m)\, d\mu =1$.
In the time interval $[t,t+dt]$ , the mass $\mu_i$ is transfered from site $i$
to site $i+1$, simultaneously for all sites $i$ (see Fig. (\ref{fig:transport})). 
In a ring geometry, one 
identifies
the site $(L+1)$ with the site $1$. Thus, after this transfer, the new masses 
at times $t+dt$ are given by~\cite{EMZ1}
\begin{equation}
m_i(t+dt)= m_i(t) -\mu_i(t) +\mu_{i-1}(t)
\label{mtr1}
\end{equation}
where the second term on the right hand side denotes the mass that has left 
site $i$
and the third term denotes the mass that came to site $i$ from site $(i-1)$.
The function $\phi(\mu|m)$ that specifies the distribution of the
stochastic mass to be transfered from any given site will be called
the `chipping kernel'. Here we take a homogeneous chipping kernel
$\phi(\mu|m)$ which does not depend on the site index. 
Note that the model above has been
defined with parallel dynamics where all sites are updated simultaneously.
Of course, by chossing $dt\to 0$, one can recover the continuous-time 
random sequential dynamics
where the probability that two sites will be updated simultaneously is very small 
$\sim O((dt)^2)$. Thus the parallel dynamics includes the continuous-time
(random sequential) dynamics as a special case. Note that for random sequential
dynamics $\phi(\mu|m)$ must generically be of the form
\begin{equation}
\phi(\mu|m) = \alpha(\mu|m) dt + \left[1-dt \int_0^m \alpha(\mu'|m)
d\mu'\right] \delta(\mu)
\label{rsd1}
\end{equation}
where $\alpha(\mu|m)$ denotes the {\it rate} at which a mass $\mu$ leaves
a site with mass $m$ and the second term denotes the probability
that no mass leaves the site. The form in Eq. (\ref{rsd1}) is designed
so that it automatically satisfies the normalization condition: $\int_0^m 
\phi(\mu|m)\, d\mu=1$.  

This model with a general chipping kernel $\phi(\mu|m)$ includes the previously 
discussed ZRP, chipping model and ARAP as special cases~\cite{EMZ1}. 
Since we introduced these models in the previous section in continuous time,
we will consider here the generalized model with chipping kernel
of the form in Eq. (\ref{rsd1}) with a general chipping rate $\alpha(\mu|m)$.
But of course one can consider a more general $\phi(\mu|m)$ with parallel
dynamics that includes the continuous-time dynamics as a special case.
Let us consider the three examples:

\vspace{0.2cm}

\noindent (i) As a first example, we see that 
ZRP is recovered if in Eq. (\ref{rsd1}) we choose for $0\le \mu\le m$
\begin{equation}
\alpha(\mu|m)= U(m)\delta(\mu-1)
\label{ckzrp}
\end{equation}
Note that $U(m)$ is zero if $m<1$.

\vspace{0.2cm}

\noindent (ii) Similarly, the asymmetric chipping model is recovered is we choose
\begin{equation}
\alpha(\mu|m)= w\delta(\mu-1)+ \delta(\mu-m) 
\label{ckchip}
\end{equation}
where the first term refers to the event of a transfer of single unit of mass with
rate $w$, the second term refers to the transfer of the full mass $m$ with
rate $1$.

\vspace{0.2cm}

\noindent (iii) Finally, one recovers ARAP by choosing
\begin{equation}
\alpha(\mu|m)= \frac{1}{m} 
\label{ckarap}
\end{equation}
for all $0\le \mu\le m$ corresponding to the transfer of a fraction
of mass that is chosen uniformly in $[0,1]$ leading to a uniform
rate $\alpha(\mu|m)$ independent of $\mu$.

By appropriately choosing the chipping kernel $\phi(\mu|m)$, or equivalently
the rate $\alpha(\mu|m)$ for continuous-time dynamics, 
one can construct a whole class of mass transport models thus
justifying the name `generalized mass transport model'. 

Given a general chipping kernel $\phi(\mu|m)$, or equivalently
the chipping rate $\alpha(\mu|m)$ for continuous-time dynamics, one can ask two important
questions: (i) What is the steady state joint mass distribution $P(m_1,m_2,\ldots,m_L)$?
(ii) Which types of $\phi(\mu|m)$, or equivalently $\alpha(\mu|m)$ for
continuous-time dynamics, lead to a real-space 
condensation transition
in the steady state? As discussed earlier, 
answers to either of these questions
are hard to provide for a general chipping kernel $\phi(\mu|m)$ (or chipping rate
$\alpha(\mu|m)$). However, 
let us now
restrict ourselves only to those chipping kernels $\phi(\mu|m )$ (or $\alpha(\mu|m)$) that 
lead to a factorised
steady state distribution of the form
\begin{equation}
P(m_1,m_2,\ldots, m_L) = \frac{1}{Z_{L}(M)} f(m_1)\,f(m_2)\ldots 
f(m_L)\,\delta\left(\sum_i m_i-M\right)
\label{ckfact1}
\end{equation}
where the partition function $Z_L(M)$ is just a normalization constant.
This then leads us to the restricted mass transport model with factorisable steady state
and we will address the questions regarding real-space condensation within this
restricted class. The answers to these questions turn out to be easier for
this restricted class since one can 
make use of the exact form
of the steady state joint mass distribution (\ref{ckfact1}).

\vspace{0.2cm}

{\bf A Restricted Mass Transport Model:} Here we restrict ourselves only
to those chipping kernels that lead to a factorisable steady state 
(\ref{ckfact1}). Let us first investigate the question: given $\phi(\mu|m)$ (or 
equivalently $\alpha(\mu|m)$ for continuous-time dynamics),
what is the necessary and sufficient condition on $\phi(\mu|m)$ that leads
to a factorisable steady state as in Eq. (\ref{ckfact1}) and if it happens, what is the exact 
form of the weight 
function $f(m)$ in terms of $\phi(\mu|m)$? Fortunately, answers to
both questions can be obtained exactly which we state below without giving the details of 
the proof (see Ref. (\cite{EMZ1}) for details). The necessary and sufficient
condition for factorisability is that $\phi(\mu|m)$ must be of the form~\cite{EMZ1}
\begin{equation}
\phi(\mu|m)= \frac{v(\mu)\, w(m-\mu)}{\int_0^m v(y)\,w(m-y)\, dy}
\label{nsc1}
\end{equation}
where $v(x)$ and $w(x)$ are arbitrary positive functions and the 
denominator is chosen to ensure normalizability, $\int_0^m \phi(\mu|m) d\mu=1$.
In other words, if the chipping kernel $\phi(\mu|m)$ factorises into
a function of the mass that leaves the site and a function of the mass 
that stays on the site, then the steady state is guaranteed to be factorisable
as in Eq. (\ref{ckfact1}) with a weight function whose exact form is
given by the denominator in Eq. (\ref{nsc1})
\begin{equation}
f(m)= \int_0^m v(y)\,w(m-y)\, dy.
\label{weight1}
\end{equation}
This is the sufficiency condition. On the other hand, this condition
can also be proved to be necessary, i.e., given that the steady state 
is of the form (\ref{ckfact1}) with some $f(m)$, the chipping kernel has to be
of the type (\ref{nsc1}) and $f(m)$ then must have the
form (\ref{weight1}).

Analogous condition can be found for continuous-time dynamics where   
$\phi(\mu|m)$ has the form (\ref{rsd1}) with a chipping
rate $\alpha(\mu|m)$. The necessary and sufficient condition stated
above for $\phi(\mu|m)$, translates into the following condition on 
$\alpha(\mu|m)$~\cite{EMZ1},
\begin{equation}
\alpha(\mu|m) = y(\mu)\, \frac{z(m-\mu)}{z(m)}
\label{nscrsd}
\end{equation}
where $y(x)$ and $z(x)$ are two arbitrary positive functions. Note that
$\alpha(\mu|m)$ is a rate (and not a probability) and hence there is
no normlization condition here to be satisfied. If the rate $\alpha(\mu|m)$ 
has the form (\ref{nscrsd}) then we are gauranteed to reach a 
factorisable steady state (\ref{ckfact1}) with a
simple weight function~\cite{EMZ1}   
\begin{equation}
f(m)= z(m).
\label{weightrsd}
\end{equation}

As an example, it is easy to verify that the chipping rate
in ZRP (\ref{ckzrp}) can indeed be written in the form
in Eq. (\ref{nscrsd}) by choosing $y(\mu)=\delta(\mu-1)$
and $z(m) = \prod_{k=1}^m \frac{1}{U(k)}$ for $m\ge 1$, $z(0)=1$
and $z(m<0)=0$. Thus, ZRP
with sequential dynamics is gauranteed to have a factorisable steady state
(\ref{ckfact1}) with the weight function,
$f(m) = z(m) = \prod_{k=1}^m \frac{1}{U(k)}$ for $m\ge 1$ 
and $f(0)=z(0)=1$. In contrast, both for asymmetric chipping model
and the ARAP with chipping rates given respectively in (\ref{ckchip})
and (\ref{ckarap}) can not be written in the form (\ref{nscrsd})
with some choice of nonnegative functions $y(x)$ and $z(x)$, proving
that neither of these two models has a factorisable steady state.

One can ask several other interesting related questions. For example,
suppose we are given a chipping kernel $\phi(\mu|m)$ and we want
to know if it has a factorisable steady state or not. This amounts 
to an explicit search for 
suitable nonnegative functions $v(x)$ and 
$w(x)$ such that the kernel $\phi(\mu|m)$ can be written
as in Eq. (\ref{nsc1}). This is often laborious. Can one devise
a simple test which will allow us to do this search quickly just by looking
at the functional form of $\phi(\mu|m)$? It turns out that indeed
one can devise such a simple test which is stated as follows~\cite{ZEM}:
Given $\phi(\mu|m)$, first set $m=\mu+\sigma$ and compute
the following two derivatives
\begin{equation}
q(\mu, \sigma)= \partial_\mu \partial_\sigma \log\left[\phi(\mu|\mu+\sigma)\right]
\label{test1}
\end{equation}
which, in general, is a function of both variables $\mu$ and $\sigma$.
The test devised in Ref. \cite{ZEM} states that a given $\phi(\mu|m)$ 
will lead to a factosibale steady state iff this function $q(\mu,\sigma)$
in Eq. (\ref{test1}) is only a function of the single variable $\mu+\sigma$, i.e.,
\begin{equation}
q(\mu,\sigma)= h(\mu+\sigma)
\label{test2}
\end{equation}
and in that case the weight function $f(m)$ in the factorisable steady state
(\ref{ckfact1}) is given explicitly by
\begin{equation}
f(m) = \exp\left[-\int^m dx \int^x dy\, h(y)\right].
\label{test3}
\end{equation}
  
In the discussion above, we have focused only on the ring geometry
with asymmetric transfer of mass. Some of these results can be partially generalized
to higher dimensions and even to arbitrary graphs~\cite{EMZgraph,GL} and also
to transport models with more than one species~\cite{Hanney1} of scalar variables, such
as mass and energy for instance.

\section{Condensation in Mass Transport Models with Factorisable Steady State} 

In this section we discuss issues related to condensation within the restricted
class of mass transport models that have a special steady state--namely a 
factorisable joint distribution (\ref{ckfact1}) with a suitable weight function 
$f(m)$. We note that some aspects of the
condensation transition in such a factorisable 
steady state, notably properties in the fluid state,
was first studied in the context of the Backgammon model~\cite{BBJ} without recourse to the
dynamics that gives rise to such a factorisable steady state.
A more complete analysis including the study of the properties of 
the condensed phase was undertaken in Refs. \cite{MEZ,EMZlong} which
will be summarized in this section.

There are three main issues here: (i) {\it criterion}: what kind of weight functions $f(m)$
lead to a condensation transition (ii) {\it mechanism}: what is the mechanism of
the condensation transition when there is one and (iii) {\it nature}: what is the
nature of the condensate, e.g., what is the statistics of the mass in the condensate
in the condensed phase? All these questions can be answered in detail
for factorisable steady states. We briefly mention the main results
here, the details can be found in Refs. \cite{MEZ,EMZlong}.

\vspace{0.2cm}

\noindent {\bf Crietrion:} The factorisation property (\ref{ckfact1})
allows one to find the criterion for a 
condensation transition rather easily by working in the grand canonical ensemble (GCE).
Within GCE framework, one introduces a fugacity $\exp[-s\, m]$ where
$s$ is the negative of the chemical potential associated with each site. 
This is just equivalent to taking the Laplace transform of Eq. (\ref{ckfact1})
with respect to the total mass $M$ (with $s$ being the Laplace variable), 
which replaces the delta 
function by $\exp[-s(m_1+m_2+\ldots m_L)]$. Then $s$ is chosen such that the constraint
$M=\sum m_i$ is satisfied on an average. Given that each site now has a mass distribution
$p(m) = f(m) \exp[-s\, m]$ (upto a normalization constant), the equation that
fixes the value of $s$ for a given $M=\rho L$ is simply
\begin{equation}
\rho = \rho(s)\equiv \frac{\int_0^\infty mf(m)e^{-s m}%
\ensuremath{\mathrm{d}}m}{\int_0^\infty f(m)e^{-s m}\ensuremath{\mathrm{d}}%
m}.  
\label{cp1}
\end{equation}
The criterion for condensation can be derived easily by analysing the
function $\rho (s )$ defined in Eq. (\ref{cp1}). 
If for a given $\rho$, one finds a solution to this equation $s=s^*$
such that the single site mass distribution is normalizable, i.e.,
 $\int p(m)\,dm=\int f(m)\,\exp[-s^*\,m]\, dm$ is finite, then there is no
condensation in the sense that for all values of $\rho$, the single site mass distribution 
has an exponential tail and there is not one special site that needs
to accomodate extra mass. 
On the other hand, it may be that for certain $f(m)$'s, as one increases
$\rho$,
there may be a critical value $\rho_c$ below which one finds
a good solution $s$ to Eq. (\ref{cp1}), but such a solution
ceases to exist for $\rho>\rho_c$. This will then signal the
onset of a condensation
because for $\rho>\rho_c$, the system needs to
break up into two parts: (a) a critical background fluid part consisting of $(L-1)$ sites 
at each of which the average density is critical $\rho_c$ and (b) a single condensate site
which accomodates the additional mass $(\rho-\rho_c)L$.  

As an example, let us consider $f(m)$ that decays slower than an exponential, but
faster than $1/{m^2}$ for large $m$.
Since $f(m)$ decays slower than an exponential, in order that the single site mass 
distribution $p(m)= f(m)\,e^{-s^*\, m}$
is normalizable (i.e., $\int p(m)\, dm =1$), the possible solution $s^*$ of Eq. (\ref{cp1}) 
can not be negative. Thus the
lowest possible solution is $s^*=0$. Now as $s\to 0$, the function $\rho(s)$
in (\ref{cp1}) approaches a critical value,
\begin{equation}
\rho_c=\rho(s\to 0)= \frac{\int_0^{\infty} m\, f(m)\, dm}{\int_0^{\infty} f(m)\, dm}
\label{rhoc}
\end{equation}
which is finite since $f(m)$ decays faster than $1/m^2$ for large $m$.
Thus as long as $\rho<\rho_c$, by solving (\ref{cp1}) one will get a positive solution $s^*$
and hence no condensation. As $\rho\to \rho_c$ from below, $s^*\to 0$ from above.
But for $\rho>\rho_c$, there is no positive solution $s^*$ to (\ref{cp1}), which
signals the onset of a condensation transition.

A detailed
analysis of Eq. (\ref{cp1}) shows~\cite{EMZlong} that in order to have condensation,
the weight function $f(m)$ must have a large $m$ tail that lies above an exponential 
but below $1/{m^2}$, i.e., $\exp[-c m]< f(m) <1/{m^2}$ for large $m$ with some positive constant
$c>0$. A natural candidate satisfying this criterion is 
\begin{equation}
f(m)\simeq A\,m^{-\gamma }\quad \mathrm{with}\quad \gamma >2  
\label{fm1}
\end{equation}  
for large $m$.
Indeed, the ZRP discussed in the previous sections with the choice
$U(m) \sim (1+\gamma/m)$ for large $m$ leads to a weight function
$f(m)$ (\ref{fm1}) and then condensation
happens only for $\gamma>2$.

\vspace{0.2cm}

\noindent {\bf Mechanism and Nature:} Given an appropriate weight function $f(m)$ 
such as in (\ref{fm1}) 
that
leads to condensation, one can then ask about the mathematical mechanism 
that drives the condensation. Actually, there is a very simple
way to understand this mechanism in terms of sums of random variables~\cite{EMZlong}
which we will discuss in the next section. For now, we notice  
that in an infinite system,
where GCE is appropriate, the single site mass distribution $p(m)$
has the form $p(m)=f(m)\exp[-sm]$ with an appropriate $s$ which is the solution of 
Eq. (\ref{cp1}) as long as $\rho<\rho_c$. For $\rho>\rho_c$, there is no
solution to Eq. (\ref{cp1}). In fact, for $\rho>\rho_c$, the value of $s$ sticks to
its critical value $s_c$ and the GCE framework is no longer valid. To understand
how the condensation manifests itself in the single site mass distribution, one has
to study a system with a finite size $L$ and work in the canonical
ensemble with a strict delta function constraint as in Eq. (\ref{ckfact1}).

In a finite system of size $L$,
the single site mass distribution $p(m)$ can be obtained by integrating
the joint distribution (\ref{ckfact1}) over the masses at
all sites except one where the mass is fixed at $m$. It is easy to
see from Eq. (\ref{ckfact1}) that
\begin{equation}
p(m)= \int P(m,m_1,m_2,\ldots m_L)dm_2\, dm_3\ldots dm_L = f(m) \frac{Z_{L-1}(M-m)}{Z_L(m)}
\label{sspm1}
\end{equation}
where the partition function $Z_L(M)$
\begin{equation}
Z_L(M)= \int f(m_1)f(m_2)f(m_L)\delta\left(\sum_i m_i -M\right) dm_1\,dm_2\ldots dm_L.
\label{pf1}
\end{equation} 
Taking Laplace transform with respect to $M$, one gets
\begin{equation}
\int_0^{\infty} Z_L(M)\, e^{-sM}\, dM = \left[\int_0^{\infty} f(m)\, e^{-sm}\, dm\right]^L
\label{lptr1}
\end{equation}
which can be formally inverted using the Bromwich formula 
\begin{equation}
Z_L(M)=\int_{s_0-i\infty }^{s_0+i\infty }\frac{\ensuremath{\mathrm{d}} s}{%
2\pi i}\exp \left[ L\left( \ln g(s)+\rho s\right) \right]  \label{brom1}
\end{equation}
where we have used $M=\rho L$, the integral runs along the imaginary axis $(s_0+i\,y)$
in the complex $s$ plane to the right of all singularities of the
integrand and  
\begin{equation}
g(s)\equiv \int_0^{\infty} f(m)\, e^{-s\,m}\, \ensuremath{\mathrm{d}} m.
\label{ltfm}
\end{equation}
One can write a similar integral representation of the numerator 
$Z_L(M-m)$ in Eq. (\ref{sspm1}).  Next one analyses $Z_L(M)$
and $Z_L(M-m)$ using the method of steepest decent in the large $L$ limit.
  
As long as there is a saddle point solution to Eq. (\ref{brom1}), say at
$s=s^*$, we see immediately from Eq. (\ref{sspm1}) that the single
site mass distribution for large $L$ has the form, $p(m)\sim f(m) \exp[-s^*m]$,
i.e., one recovers the GCE result. Thus the GCE aprroach is valid as
long as there is a saddle point solution $s^*$. As one increases $\rho$, the
saddle $s^*$ starts moving towards $0$ in the complex $s$ plane and when $\rho$
hits $\rho_c$, $s^*\to 0$. For $\rho>\rho_c$, there is no saddle point
and one has to analyse the Bromwich integrals by correctly choosing the
contour to evaluate $p(m)$ for $\rho>\rho_c$. This was done in details
in Ref. \cite{EMZlong}. 
We omit the details here and mention the main
results for $Z_L(M)$ in Eq. (\ref{brom1}) and subsequently for $p(m)$ in Eq.
(\ref{sspm1}), when the weight function 
$f(m)$ is 
chosen to
be of the power law form (\ref{fm1}).

If we normalize $f(m)$ such that $\int_0^{\infty} f(m)\,dm=1$, 
the partition function in Eq. (\ref{brom1}) can be interpreted
as the probability that the sum of $L$ i.i.d variables, each drawn from
a distribution $f(m)$, is $M$ (see Section V also). Analysing the
Bromwich integral (\ref{brom1}) for large $L$, using the small $s$
behavior of $g(s)$ in Eq. (\ref{ltfm}), one finds~\cite{EMZlong}
that the asymptotic behavior of the distribution $Z_L(M)$
is different for $2<\gamma\le 3$ and for $\gamma>3$.

\vspace{0.3cm}

\noindent { $\bf {2<\gamma\le 3}$:} In this regime, one finds the
following scaling behavior of $Z_L(M)$
\begin{equation}
Z_L(M) \simeq \frac{1}{L^{1/(\gamma-1)}}\, V_{\gamma}\left[ 
\frac{\rho_c L-M}{L^{1/(\gamma -1)}}\right]
\label{pfsb_an}
\end{equation}
where $\rho_c= \mu_1=\int_0^{\infty} m\, f(m)\,dm$ is the first moment and the function 
$V_\gamma (z)$ is given 
explicitly by~\cite{EMZlong}
\begin{equation}
V_{\gamma}(z)=\frac{1}{\pi} \int_0^{\infty} \ensuremath{\mathrm{d}} y\,
e^{-c_3 y^{\gamma-1}} \cos\left[b\cos(\pi\gamma/2)y^{\gamma-1}+yz\right].
\label{scf2}
\end{equation}
Here $c_3= -b\sin(\pi\gamma/2)>0 $ and $b= A\, \Gamma(1-\gamma)$
for $2<\gamma <3$ with $A$ being the amplitude in (\ref{fm1}).
The precise asymptotic tails of this scaling
function can
be computed~\cite{EMZlong}
\begin{eqnarray}
V_\gamma (z) &\simeq &A\,|z|^{-\gamma }\quad \mathrm{as}\,\,z\to -\infty
\label{vzn} \\
&=& c_0 \quad \mathrm{at}\,\, z=0  \label{vz0} \\
&\simeq & c_1\, z^{(3-\gamma)/{2(\gamma-2)}}\, e^{-c_2
z^{(\gamma-1)/(\gamma-2)}} \quad \mathrm{as}\,\, z\to \infty  \label{vzp}
\end{eqnarray}
where $c_0$, $c_1$ and $c_2$ are known constants~\cite{EMZlong}.
Thus the function is manifestly non-gaussian.

\vspace{0.3cm}

\noindent { $\bf {\gamma>3}$:} In this regime, the partition function 
$Z_L(M)$ has a gaussian peak 
\begin{equation}
Z_L(M)\simeq \frac {1}{\sqrt{2\pi \Delta ^2\,L}}
\,e^{-(M-\rho_c L)^2/{2\Delta ^2L}}\quad \mathrm{for}\,\,|M-\rho_c L|\ll
O(L^{2/3})
\label{pfsb_g}
\end{equation}
where  $\Delta^2 = \mu_2-\mu_1^2$ with
$\mu_k= \int_0^{\infty} m^k\, f(m)\, dm$ being the $k$-th moment.
But far to the left of the peak, $Z_L(M)$ has a power law decay~\cite{EMZlong}.

\vspace{0.2cm}

\begin{figure}[tbph]
\includegraphics[height=10.0cm,width=10.0cm,angle=0]{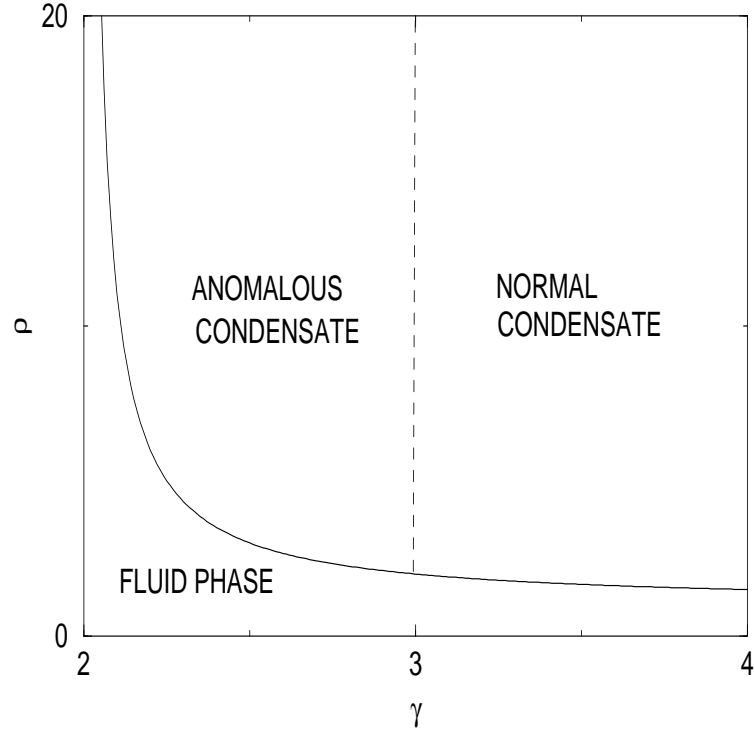}
\caption{Schematic phase diagram in the $\rho $--$\gamma $ plane.}
\label{fig:phd}
\end{figure}

So, how does the single site mass distribution $p(m)$ in Eq. (\ref{sspm1})
look like? We have to use the result for the partition function derived
above in Eq. (\ref{sspm1}). We find 
different behaviors of $p(m)$ in different regions of the $(\rho -\gamma )$
plane. For $\gamma >2$, there is a critical curve $\rho
_c(\gamma )$ in the $(\rho -\gamma )$ plane that separates a fluid phase
(for $\rho <\rho _c(\gamma )$) from a condensed phase (for $\rho >\rho
_c(\gamma )$). In the fluid phase the mass distribution decays exponentially
for large $m$, $p(m)\sim \exp [-m/m^{*}]$ where the characteristic mass $%
m^{*}$ increases with increasing density and diverges as the density
approaches its critical value $\rho _c$ from below. At $\rho =\rho _c$ the
distribution decays as a power law, $p(m)\sim m^{-\gamma }$ for large $m$.
For $\rho >\rho _c$, the distribution, in addition to the power law decaying
part, develops an additional bump, representing the condensate, centred
around the ``excess'' mass:
\begin{equation}
M_{ex}\equiv M-\rho _c\,L.  \label{Mexcess}
\end{equation}
Furthermore, by our analysis within the canonical ensemble, we show that
even inside the condensed phase ($\rho >\rho _c(\gamma )$), there are two
types of behaviors of the condensate depending on the value of $\gamma $.
For $2<\gamma <3$, the condensate is characterized by anomalous non-gaussian
fluctuations whereas for $\gamma >3$, the condensate has gaussian
fluctuations. This leads to a rich phase diagram in the $(\rho -\gamma )$
plane, a schematic picture of which is presented in Fig. (\ref{fig:phd}).

\begin{figure}[tbph]
\includegraphics[height=10.0cm,width=10.0cm,angle=0]{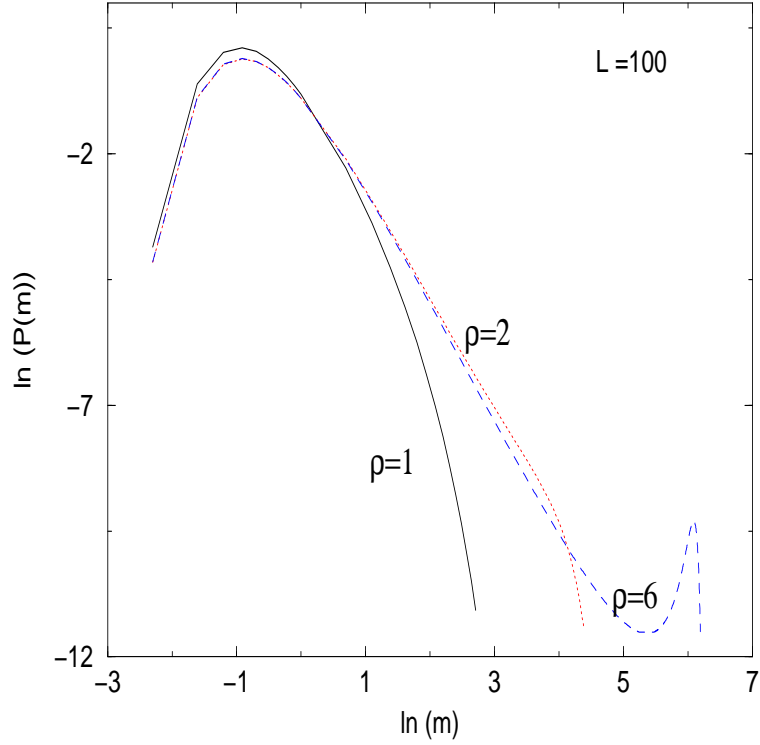}
\caption{The exact single-site mass distribution $p(m)$ plotted for
a particular choice of $f(m)$ with $\gamma=5/2$ and $\rho_c=2$, and system size $L=100$:
full line $\rho=1$ (subcritical: fluid phase); dotted line $\rho=2$ (critical); sadhes line $\rho=6$
(supercritical: condensed phase). the condensate bump, $p_{\rm cond}(m)$, is evident in the
supercritical phase.}
\label{fig:dist}
\end{figure}

Detailed form of $p(m)$ for $\rho<\rho_c$ (fluid phase), $\rho>\rho_c$ (condensed phase)
and $\rho=\rho_c$ (critical point) are summarized below (see Fig. (\ref{fig:dist}) also).

\vspace{0.2cm}

\noindent \textbf{Fluid phase} $\rho<\rho_c$: In this case one finds 
\begin{equation}
p(m)\sim f(m) \, e^{-m/m^*}\quad\mbox{for}\quad 1\ll m \ll M
\end{equation}
where the characteristic mass $m^*$ diverges $\rho$ approaches $\rho_c$ from
below as $(\rho-\rho_c)^{-1}$ for $\gamma >3$ and $(\rho-\rho_c)^{-1/(%
\gamma-2)}$ for $2<\gamma <3$.\\

\noindent \textbf{Condensed Phase} $\rho >\rho _c$: In this case one finds
\begin{eqnarray}
p(m) &\simeq &f(m)\quad \mbox{for}\quad 1\ll m\ll O(L) \\
p(m) &\simeq &f(m)\frac 1{(1-x)^\gamma }\quad \mbox{for}\quad m=xM_{ex}\quad %
\mbox{where}\quad 0<x<1 \\
p(m) &\sim &p_{\mathrm{cond}}(m)\quad \mbox{for}\quad m\sim M_{ex}
\end{eqnarray}
Here $p_{\mathrm{cond}}$ is the piece of $p(m)$ which describes the
condensate bump (see Fig. (\ref{fig:dist}): Centred on the excess $M_{ex}$ and with integral being equal to 
$%
1/L$, it takes on two distinct forms 
according to whether $\gamma <3$ or $\gamma >3$. 

For $2<\gamma <3$
\begin{equation}
p_{\mathrm{cond}}(m)\simeq L^{-\gamma /(\gamma -1)}\,V_\gamma \left[ \frac{%
m-M_{ex}}{L^{1/(\gamma -1)}}\right] ,
\label{pcond1}
\end{equation}
where the function $V_\gamma (z)$ is given explicitly 
in (\ref{scf2}).
Thus clearly the condensate bump has a non-gaussian shape
for $2<\gamma <3$ and we refer to this as an `anomalous'
condensate. 

On the other hand, for $\gamma >3$
\begin{equation}
p_{\mathrm{cond}}(m)\simeq \frac 1{\sqrt{2\pi \Delta ^2L^3}%
}\,e^{-(m-M_{ex})^2/{2\Delta ^2L}}\quad \mathrm{for}\,\,|m-M_{ex}|\ll
O(L^{2/3}).
\label{pcond2}
\end{equation}
i.e. $p_{\mathrm{cond}}(m)$ is gaussian on the scale $|m-M_{ex}|\ll
O(L^{2/3})$, but, far to the left of the peak, $p(m)$ decays as a power law.\\

\noindent \textbf{Critical density} $\rho=\rho_c$: In this case one finds
that
\begin{eqnarray}
p(m)&\propto& f(m) V_{\gamma}\left(m/L^{1/(\gamma-1)}\right), \quad\mbox{for}%
\quad 2<\gamma<3 \\
p(m)&\propto& f(m) \, e^{-m^2/{2\Delta^2 L}}\quad \gamma>3 .  \label{m2}
\end{eqnarray}
where the scaling function $V_{\gamma}(z)$ is as before. Thus at
criticality $p(m)$ decays as a power law $m^{-\gamma}$ for large $m$ which
is cut-off by a finite size scaling function and the cut-off mass scales as
\begin{eqnarray}
m_{\mathrm{cut-off}} &\sim & L^{1/(\gamma-1)} \quad\quad \mathrm{for}\,\,\,
2<\gamma<3  \label{m1} \\
&\sim & L^{1/2} \quad\quad \mathrm{for}\,\,\, \gamma>3 .
\end{eqnarray}

\vspace{0.2cm}

\noindent{\bf Physical picture:} It is useful to summarize the main physical picture that
emerges out of this mathematical analysis. We notice from the joint distribution 
(\ref{ckfact1}) that the masses at each site are `almost' independent random variables each
with a power law distribution $f(m)$, except for the global constraint of mass conservation
imposed by the delta function which actually makes them `correlated'. The system
feels this correlation for $\rho<\rho_c$ and $\rho>\rho_c$ in different ways
and exactly at the critical point $\rho=\rho_c$ the effect of the
constraint is actually the least. For $\rho<\rho_c$, the effective
mass distribution at each site acquires an exponential tail, $p(m)\sim f(m)\exp[-s^* 
m]$ which is induced by the constraint. For $\rho=\rho_c$, $s^*\to 0$ and
$p(m)\sim f(m)$, thus the system does not feel the constraint at all and
the masses behave as completely independent random variables each distributed via $f(m)$.
But for $\rho>\rho_c$, while $(L-1)$ sites behave as the critical fluid, i.e.,
as if the mass at each of these $(L-1)$ sites is distributed
via $f(m)$, there is one single condensate site which acquires 
the additional mass $(\rho-\rho_c)L$.
 
For $\rho>\rho_c$, the resulting non-monotonous shape of the single site mass distribution (with
an additional bump) in Fig. (\ref{fig:dist}) can then be understood very easily
from this physical picture. Basically, for $\rho>\rho_c$, the total mass $M$
of the system splits into the critical fluid and the condensate part:
\begin{equation}
M= m_{\rm cond} + M_{\rm fluid}
\label{massdiv1}
\end{equation}
where $m_{\rm cond}$ denotes the mass at the condensate and the critical
background fluid mass
\begin{equation}
M_{\rm fluid} = \sum_{i=1}^{L-1} m_i
\label{fluid1}
\end{equation}
is a sum of $(L-1)$ independent random variables (masses) each distributed via
$f(m)$. Thus the probability distribution $M_{\rm fluid}$ is given precisely by the 
partition function $Z_{L-1}\left(M_{\rm fluid}\right)$ in Eq. (\ref{pf1}). This partition
function can be computed explicitly and the results are given in Eqs. (\ref{pfsb_an})
and (\ref{pfsb_g}).
Knowing this partition function, 
the distribution of the condensate mass $m_{\rm cond}$ can be
obtained using Eq. (\ref{massdiv1}) giving
\begin{equation}
{\rm Prob}(m_{\rm cond}=y)= Z_{L-1}(M-y).
\label{condmass_dist}
\end{equation}
The overall single site mass distribution for $\rho>\rho_c$ then can be computed
as follows: if we choose a site at random, with probability $(L-1)/L$ it belongs
to the background fluid and hence its mass distribution is $f(m)$ whereas
with probability $1/L$ it will be the condensate site with mass distribution
given in Eq. (\ref{condmass_dist}). Thus for $\rho>\rho_c$
\begin{equation}
p(m) \approx \frac{(L-1)}{L}\, f(m) + \frac{1}{L}\, Z_{L-1}(M-m)
\label{pmfinal}
\end{equation}
The second term is what we referred to before as 
\begin{equation}
p_{\rm cond}(m)=\frac{1}{L}\, Z_{L-1}(M-m)
\label{pcondinter}
\end{equation}  
and it is this piece that describes the bump in Fig. (\ref{fig:dist}) for $\rho>\rho_c$.
Its precise asymptotic behavior is detailed in Eqs. (\ref{pcond1}) and (\ref{pcond2})
respectively for $2<\gamma\le 3$ and $\gamma>3$.

\section{Intepretation as sums and extremes of random variables}

There is a very nice
and simple way~\cite{EMZlong}, using sums of random variables, to understand the mechanism of 
the 
condensation transition
for factorsable steady states (\ref{ckfact1}) with a given weight function $f(m)$
say of the form (\ref{fm1}). 
Let us consider a set of $L$ positive i.i.d
random variables $\{m_1,m_2,\ldots, m_L\}$ each drawn from a distribution $f(m)$
(we choose $f(m)$ such that it is normalized to unity). 
Let $M=\sum_{i=1}^L  m_i$ be the sum. For instance, $M$ can be interpreted
as the position of a random walker after $L$ independent steps of lengths
$m_1$, $m_2$, $\ldots$, $m_L$. Then we notice that the partition function
$Z_L(M)$ in Eq. (\ref{pf1}) can be interpreted as the probability
that the walker reaches $M$ in $L$ steps starting from the origin.

How does one interpret condensation in this random walk language?
Note from Eq. (\ref{rhoc}) that
for normlaized (to unity) $f(m)$, the critical density
\begin{equation}
\rho_c= \mu_1 = \int_0^{\infty} m \, f(m)\, dm
\label{rhoc1}
\end{equation}
is just the mean step length of the random walker's steps. Thus
if the final position $M< \rho_c L = \mu_1 L$ (i.e., $\rho=M/L<\rho_c$), we would expect
that the typical configuration of the random walker's path would
consist of steps each of which is of $\sim O(1)$. But, for $M> \mu_1 L$ (i.e., 
$\rho>\rho_c$),
the ensemble will be dominated by configurations where $(L-1)$ steps are of
$\sim O(1)$, but one single big step of $\sim (M-\mu_1 (L-1))\sim (\rho-\rho_c)L$
to compensate for the deficit distance. This single big step is precisely
the condensate. Within this interpretation, it also becomes clear
that for $f(m)$ of the form (\ref{fm1}), there are two possibilities
depending on whether the second moment of the step length
distribution $\mu_2= \int_0^{\infty} m^2\, f(m)\, dm$
is divergent ($2<\gamma\le 3$) or finite ($\gamma>3$). In the former case,
the corresponding random walker is a L\'evy flight with analomalosly large fluctuations
that leads to an anomalously large fluctuation in the condensate mass. In the latter case,
by virtue of the central limit theorem, one recovers a gaussian fluctuation 
leading to a gaussian distribution of the condensate mass. This explains
the two types of condensate phases in the phase diagram in Fig. (\ref{fig:phd}).

\vspace{0.2cm}

\noindent {\bf Condensation and Extreme Statistics:} Another interesting issue intimately related
to the condensation is the associated extreme value statistics--e.g, what is the distribution of the largest
mass $m_{\rm max}$ in the system, where
\begin{equation}
m_{\rm max}= {\rm max}(m_1,m_2,\ldots,m_L).
\label{mmax0}
\end{equation}
This is particularly important in the condensed phase where the 
largest
mass is carried by the condensate, at least in models with simple factorised steady states
where there is a single condensate. The theory of extreme value statistics is well developed
in cases where one studies the extreme (e.g., the maximum) of a set of 
i.i.d random variables~\cite{Gumbel}. In our case, the factorised steady state
in Eq. (\ref{ckfact1}) shows that the masses are not completely independent, but are correlated
via the global mass conservation constraint explicitly manifest in the delta function
in Eq. (\ref{ckfact1}). Without this delta function constraint, with $f(m)\sim m^{-\gamma}$
and $\gamma>2$, the scaled distribution of the maximal mass would have been Fr\'echet 
distribution~\cite{Gumbel}. However the presence of the constraint induces important correlations that
changes the nature of the distribution of the maximal mass. 

In Ref. \cite{JMP}, the authors studied rigorously how the typical value of the
extremal mass scales with system size in ZRP with $\gamma>3$. For more general
mass transport models with factorised steady states as in Eq. (\ref{ckfact1}), the
full distribution of extremal mass was studied recently in Ref. \cite{extreme}.
In was found that
in the fluid phase
($\rho<\rho_c$), where the single site masses effectively become uncorrelated but with an additional
$\exp[-s^* m]$ factor that comes from the conservation constraint: $p(m)\sim f(m)\exp[-s^*m]$   
and hence the maximal mass, in the scaling limit, becomes Gumbel~\cite{extreme}. At the critical point 
where $s^*=0$ (i.e., where the constraint is least effective), one recovers the Fr\'echet distribution.
But for $\rho>\rho_c$ the maximal mass distribution is the same as that of the condensate mass,
i.e.,
\begin{equation}
m_{\rm max}= m_{\rm cond}
\label{maxm1}
\end{equation}
However, as mentioned in the previous section, the distribution of $m_{\rm cond}= M - M_{\rm fluid}$ can be 
computed via
computing the distribution of $M_{\rm fluid}$ as a sum of independent random variables each distributed via 
$f(m)$ and  is given by Eq. (\ref{condmass_dist}). Thus,
\begin{equation}
{\rm Prob}\left(m_{\rm max}=y\right)\approx Z_{L-1}(M-y)= L\, p_{\rm cond}(y)
\label{maxm2}
\end{equation}
where  
$p_{\rm cond}(m)$ is given in Eqs. (\ref{pcond1}) and (\ref{pcond2}) respectively for
$2<\gamma<3$ and $\gamma>3$. Thus, in the condensed phase, one has a completely new type 
of extreme value distribution of correlated random variables which is exactly computable~\cite{extreme}.
Moreover, it is interesting to note that for $\rho>\rho_c$, the computation 
of the distribution of the extreme 
of correlated variables reduces to the calculation of the distribution of sum
of independent random variables.

\section{Conclusion}

In this brief review I discussed recent developments in understanding the physics
of the real-space condensation in a class of mass transport models. Here, real-space
condensation refers to the phenomenon when, upon increasing the density beyond
a critical value, a macroscopically large mass settles onto a single site in
real space in the steady state. The system discussed is homogeneous in the sense
that the transport rules do not depend on the sites or particles. Thus, 
in the limit of an infinite system size, the formation
of the condensate at one single site actually breaks the translational symmetry in the
system spontaneously. The criterion and mechanism of the transition
as well as the detailed finite size dependence of the distribution of the
condensate mass in the condensed phase was discussed within a restricted
class of one dimensional mass transport models that have a factorisable steady 
state.

There are several directions in which the questions addressed here can be 
extended, 
some of which are briefly mentioned below.

\vspace{0.2cm}

\noindent {\bf Pair factorised steady states:} Here we discussed only mass transport models 
that
have a factorised steady state (\ref{ckfact1}). A natural
question is whether real-space condensation can happen in other types of steady states 
that are not simply factorisable as in (\ref{ckfact1}) and if so, (i)
are there natural {\it local} transport rules that lead to such steady states
and (ii) does the nature of the condensate change fundamentally from the
one with factorisable steady states? Recently, a generalization of Eq. (\ref{ckfact1}),
called the pair factorised steady states (PFSS) was introduced in Ref. \cite{EHM}
\begin{equation} 
P(m_1,m_2,\ldots,m_L) ] = \frac{1}{Z_{L}(M)} \prod_{i=1}^L g(m_i, m_{i{+}1})
\,\delta\left(\sum_{i=1}^L m_i - M\right) \;.
\label{P(C)}
\end{equation}
Thus there is one factor $g(m_i, m_{i{+}1})$ for each pair of
neighbouring sites on a ring of $L$ sites.
The transport rules, involving three neighbouring sites, that lead to such steady states were also 
found~\cite{EHM}. Interestingly, the condensate in this PFSS, for a class of weight functions
$g(m,n)$ that are short ranged, was found to spread over a relatively large 
number of sites $\sim O(L^{1/2})$~\cite{EHM},
in contrast to the condensate that forms over a single site in the usual
factorised steady state (\ref{ckfact1}).  
The average shape of this sub-extensive condensate for a class of weight functions $g(m,n)$ 
as well as the precise form of the `condensate bump' in the single site mass distribution
in the condensed phase were recently computed in a very nice paper~\cite{WSJM1}.
In addition, the transport rules that lead to PFSS on an arbitrry graph
were also found recently~\cite{WSJM2}, thus generalizing the 
one dimensional models with PFSS.

\vspace{0.2cm}

\noindent {\bf Dynamics:} Here we have discussed only static properties
associated with the condensation transition. Another interesting issue is the 
nature of the dynamics in the steady state as well as in the approach to
the steady state, in particular in the condensed phase~\cite{Godreche,GSS,GL1,GBM,SH}. In a finite system,
the condensate forms at a site, then survives there over a long time $T_s$,
then desolves and then forms at another site. For ZRP with $f(m)\sim m^{-\gamma}$
with $\gamma>2$ and in the stationary state, it was found
that while the condensate life-time $T_s\sim (\rho-\rho_c)^{\gamma+1}\,L^{\gamma}$ 
for large $L$~\cite{GL1}, there is another shorter time scale associated with the 
relocation of the condensate, $T_r\sim (\rho-\rho_c)^2 L^2$~\cite{GBM}. In addition,
the current fluctuations in the steady state show a striking change
of behavior~\cite{GBM} as one goes from the fluid phase ($\rho<\rho_c$) to the
condensed side ($\rho>\rho_c$).  
On the fluid side, the current fluctuations shows an interesting oscillating behavior due to 
the presence of kinematic waves~\cite{GBM}, similar to the oscillatory behavior
of the variance in the displacement of a tagged particle in $1$-d asymmetric exclusion 
processs~\cite{GMGB}. Another interesting dynamical quantity is the power spectra associated
with the time series depicting the evolution of the total number of particles over a fixed segment
of a ring, studied recently in the context of ZRP~\cite{AZ}. 

It would be interesting to study the dynamics for the mass transport models in higher
dimensions or arbitrary graphs, and also for more generalized steady states such as 
the PFSS.

\vspace{0.2cm}

Other interesting directions involve studying the condensation phenomena in multispecies
models~\cite{zrp_review,Hanney1}, misanthropic process~\cite{zrp_review},
instability of the condensed phase due to nonconserving rates in the chipping model~\cite{MKB,MKB1}
as well as in ZRP~\cite{AELM,GS1} and also in ZRP due to quench disordered 
particle transfer rate~\cite{GCS},
ZRP leading to multiple condensates~\cite{SEM}, condensation in polydisperse hard spheres~\cite{EMTP} etc. 
Most of 
these
generalizations have been carried out so far in the context of ZRP, but it would be
interesting to study the general mass transport model discussed here with
these additional generalizations.
It would also be interesting to compute the distribution of maximal mass in PFSS where the condensate
is sub-extensive. In this case, the extremal mass is not the total mass carried by the full
condensate, but rather the site inside the condensate that carries the largest mass.

\vspace{0.2cm}

\noindent{\bf Acknowledgements:} It is a pleasure to thank my collaborators in this 
subject: M. Barma, M.R. Evans, C. Godr\`eche, S. Gupta, T. Hanney, S. Krishnamurthy, R. Rajesh,
E. Trizac and R.K.P. Zia.
I also thank A. Comtet, D. Dhar, J. Krug, K. Mallick, H. Meyer-Ortmanns, D. Mukamel, 
G. Sch\"utz and C. Sire for many useful discussions. This article is based
on a series of lectures that I first gave at the summer school 
``The Principles of the Dynamics of Non-Equilibrium Systems" held at 
the Issac Newton Institute, Cambridge (2006) and later at 
the summer school ``Exact Methods in Low-dimensional Statistical Physics and Quantum 
Computing" held at Les Houches (2008). I thank the organizers
of both the schools for hospitality. The support from Grant No. 3404-2 of ``Indo-French
Center for the Promotion of Advanced Research (IFCPAR/CEFIPRA)" is also
gratefully acknowledged.

\end{document}